\def\year{2022}\relax
\documentclass[letterpaper]{article} 

\usepackage[utf8]{inputenc}
\usepackage[T1]{fontenc}
\usepackage{hyphenat}
\usepackage{xspace}
\usepackage{amsmath}
\usepackage{amsfonts}
\usepackage{booktabs}
\usepackage{multirow}
\usepackage{makecell}
\usepackage{minibox}
\usepackage{bbm}
\usepackage{balance}
\usepackage{mathtools}
\usepackage{color}
\usepackage{marvosym}
\usepackage{ifthen}
\usepackage{textcomp}
\usepackage{enumitem}
\usepackage{verbatim}
\usepackage{algorithm}
\usepackage{algorithmic}
\usepackage{numprint}
\usepackage{balance}

\usepackage{amsthm}
\theoremstyle{plain}

\newcommand{\chatoDisplayMode}[1]{#1}

\definecolor{MyRed}{rgb}{0.6,0.0,0.0} 
\definecolor{MyBlack}{rgb}{0.1,0.1,0.1} 
\newcommand{\inred}[1]{{\color{MyRed}\sf\textbf{\textsc{#1}}}}
\newcommand{\frameit}[2]{
  \begin{center}
  {\color{MyRed}
  \framebox[.9\columnwidth][l]{
    \begin{minipage}{.85\columnwidth}
    \inred{#1}: {\sf\color{MyBlack}#2}
    \end{minipage}
  }\\
  }
  \end{center}
}

\newcommand{\note}[2][]{\chatoDisplayMode{\def\@tmpsig{#1}\frameit{{\Pointinghand} Note}{#2\ifx \@tmpsig \@empty \else \mbox{ --\em #1}\fi}}}
\newcommand{\todo}[2][]{\chatoDisplayMode{\def\@tmpsig{#1}\frameit{{\Writinghand} To-do}{#2\ifx \@tmpsig \@empty \else \mbox{ --\em #1}\fi}}}

\newcommand{\abbrevStyle}[1]{#1}

\newcommand{\ie}{\abbrevStyle{i.e.}\xspace}
\newcommand{\eg}{\abbrevStyle{e.g.}\xspace}
\newcommand{\cf}{\abbrevStyle{cf.}\xspace}

\newcommand{\vs}{\abbrevStyle{vs.}\xspace}
\newcommand{\etc}{\abbrevStyle{etc.}\xspace}

\newcommand{\Secref}[1]{Sec.~\ref{#1}}

\newcommand{\Tabref}[1]{Table~\ref{#1}}
\newcommand{\Figref}[1]{Fig.~\ref{#1}}

\newcommand{\Appref}[1]{Appendix~\ref{#1}}

\newcommand{\xhdr}[1]{\vspace{1.7mm}\noindent{{\bf #1.}}}

\newcommand{\xhdrNoPeriod}[1]{\vspace{1.7mm}\noindent{{\bf #1}}}

\newcommand{\textcite}[1]{\citeauthor{#1} \shortcite{#1}}

\newcommand{\hide}[1]{}

\makeatletter
\newcommand{\iffont}[2]{\ifthenelse{\equal{\f@family}{#1}}{#2}{}}
\makeatother

\iffont{ptm}{
  \usepackage{mathptmx}

  \DeclareSymbolFont{greek}{OML}{cmm}{m}{n}
  \DeclareMathSymbol{\alpha}{\mathalpha}{greek}{"0B}
  \DeclareMathSymbol{\beta}{\mathalpha}{greek}{"0C}
  \DeclareMathSymbol{\gamma}{\mathalpha}{greek}{"0D}
  \DeclareMathSymbol{\delta}{\mathalpha}{greek}{"0E}
  \DeclareMathSymbol{\epsilon}{\mathalpha}{greek}{"0F}
  \DeclareMathSymbol{\zeta}{\mathalpha}{greek}{"10}
  \DeclareMathSymbol{\eta}{\mathalpha}{greek}{"11}
  \DeclareMathSymbol{\theta}{\mathalpha}{greek}{"12}
  \DeclareMathSymbol{\iota}{\mathalpha}{greek}{"13}
  \DeclareMathSymbol{\kappa}{\mathalpha}{greek}{"14}
  \DeclareMathSymbol{\lambda}{\mathalpha}{greek}{"15}
  \DeclareMathSymbol{\mu}{\mathalpha}{greek}{"16}
  \DeclareMathSymbol{\nu}{\mathalpha}{greek}{"17}
  \DeclareMathSymbol{\xi}{\mathalpha}{greek}{"18}
  \DeclareMathSymbol{\pi}{\mathalpha}{greek}{"19}
  \DeclareMathSymbol{\rho}{\mathalpha}{greek}{"1A}
  \DeclareMathSymbol{\sigma}{\mathalpha}{greek}{"1B}
  \DeclareMathSymbol{\tau}{\mathalpha}{greek}{"1C}
  \DeclareMathSymbol{\upsilon}{\mathalpha}{greek}{"1D}
  \DeclareMathSymbol{\phi}{\mathalpha}{greek}{"1E}
  \DeclareMathSymbol{\chi}{\mathalpha}{greek}{"1F}
  \DeclareMathSymbol{\psi}{\mathalpha}{greek}{"20}
  \DeclareMathSymbol{\omega}{\mathalpha}{greek}{"21}
  \DeclareMathSymbol{\varepsilon}{\mathalpha}{greek}{"22}
  \DeclareMathSymbol{\vartheta}{\mathalpha}{greek}{"23}
  \DeclareMathSymbol{\varpi}{\mathalpha}{greek}{"24}
  \DeclareMathSymbol{\varrho}{\mathalpha}{greek}{"25}
  \DeclareMathSymbol{\varsigma}{\mathalpha}{greek}{"26}
  \DeclareMathSymbol{\varphi}{\mathalpha}{greek}{"27}
  \DeclareSymbolFont{otone}{OT1}{cmr}{m}{n}
  \DeclareMathSymbol{\Gamma}{\mathalpha}{otone}{0}
  \DeclareMathSymbol{\Delta}{\mathalpha}{otone}{1}
  \DeclareMathSymbol{\Theta}{\mathalpha}{otone}{2}
  \DeclareMathSymbol{\Lambda}{\mathalpha}{otone}{3}
  \DeclareMathSymbol{\Xi}{\mathalpha}{otone}{4}
  \DeclareMathSymbol{\Pi}{\mathalpha}{otone}{5}
  \DeclareMathSymbol{\Sigma}{\mathalpha}{otone}{6}
  \DeclareMathSymbol{\Upsilon}{\mathalpha}{otone}{7}
  \DeclareMathSymbol{\Phi}{\mathalpha}{otone}{8}
  \DeclareMathSymbol{\Psi}{\mathalpha}{otone}{9}
  \DeclareMathSymbol{\Omega}{\mathalpha}{otone}{10}
  \DeclareSymbolFont{syms}{OML}{cmm}{m}{it}
  \DeclareMathSymbol{\partial}{\mathord}{syms}{"40}
  \DeclareMathAlphabet{\mathbold}{OML}{cmm}{b}{it}
  \DeclareSymbolFont{largesymbols}{OMX}{cmex}{m}{n}

}
\usepackage{placeins}
\usepackage{subcaption}
\usepackage{amsmath}
\usepackage{bm}
\usepackage{marginnote}
\usepackage{aaai22}  
\usepackage{times}  
\usepackage{helvet}  
\usepackage{courier}  
\usepackage[hyphens]{url}  
\usepackage{graphicx} 
\usepackage{natbib}  
\usepackage{caption} 
\DeclareCaptionStyle{ruled}{labelfont=normalfont,labelsep=colon,strut=off} 
\usepackage{algorithm}
\usepackage{algorithmic}

\usepackage{newfloat}
\usepackage{listings}
\floatstyle{ruled}
\newfloat{listing}{tb}{lst}{}
\floatname{listing}{Listing}
\usepackage{mathptmx}

\newif\ifshowcomments
\showcommentstrue
\ifshowcomments

\else

\fi

\title{Post Approvals in Online Communities}
\author{
Manoel Horta Ribeiro,\textsuperscript{\rm 1}\thanks{Work done mostly while interning at Meta.}
Justin Cheng,\textsuperscript{\rm 2}
Robert West\textsuperscript{\rm 1}\\
}
\affiliations{
\textsuperscript{\rm 1}EPFL, \textsuperscript{\rm 2}Meta \\
manoel.hortaribeiro@epfl.ch, jcheng@fb.com, robert.west@epfl.ch
}

\begin{document}

\maketitle

\begin{abstract}
In many online communities, community leaders (\ie moderators and administrators) can proactively filter undesired content by requiring posts to be approved before publication.
But although many communities adopt post approvals, there has been little research on its impact on community behavior.
Through a longitudinal analysis of 233,402 Facebook Groups, we examined
1)~the factors that led to a community adopting post approvals and
2)~how the setting shaped subsequent user activity and moderation in the group.
We find that communities that adopted post approvals tended to do so following sudden increases in user activity (\eg comments) and moderation (\eg reported posts).
This adoption of post approvals led to fewer but higher-quality posts.
Though fewer posts were shared after adoption, not only did community members write more comments, use more reactions, and spend more time on the posts that were shared, they also reported these posts less.
Further, post approvals did not significantly increase the average time leaders spent in the group, though groups that enabled the setting tended to appoint more leaders.
Last, the impact of post approvals varied with both group size and how the setting was used, \eg, group size mediates whether leaders spent more or less time in the group following the adoption of the setting.
Our findings suggest ways that proactive content moderation may be improved to better support online communities.
\end{abstract}

\section{Introduction}

\begin{figure}
    \centering
    \includegraphics[width=\linewidth]{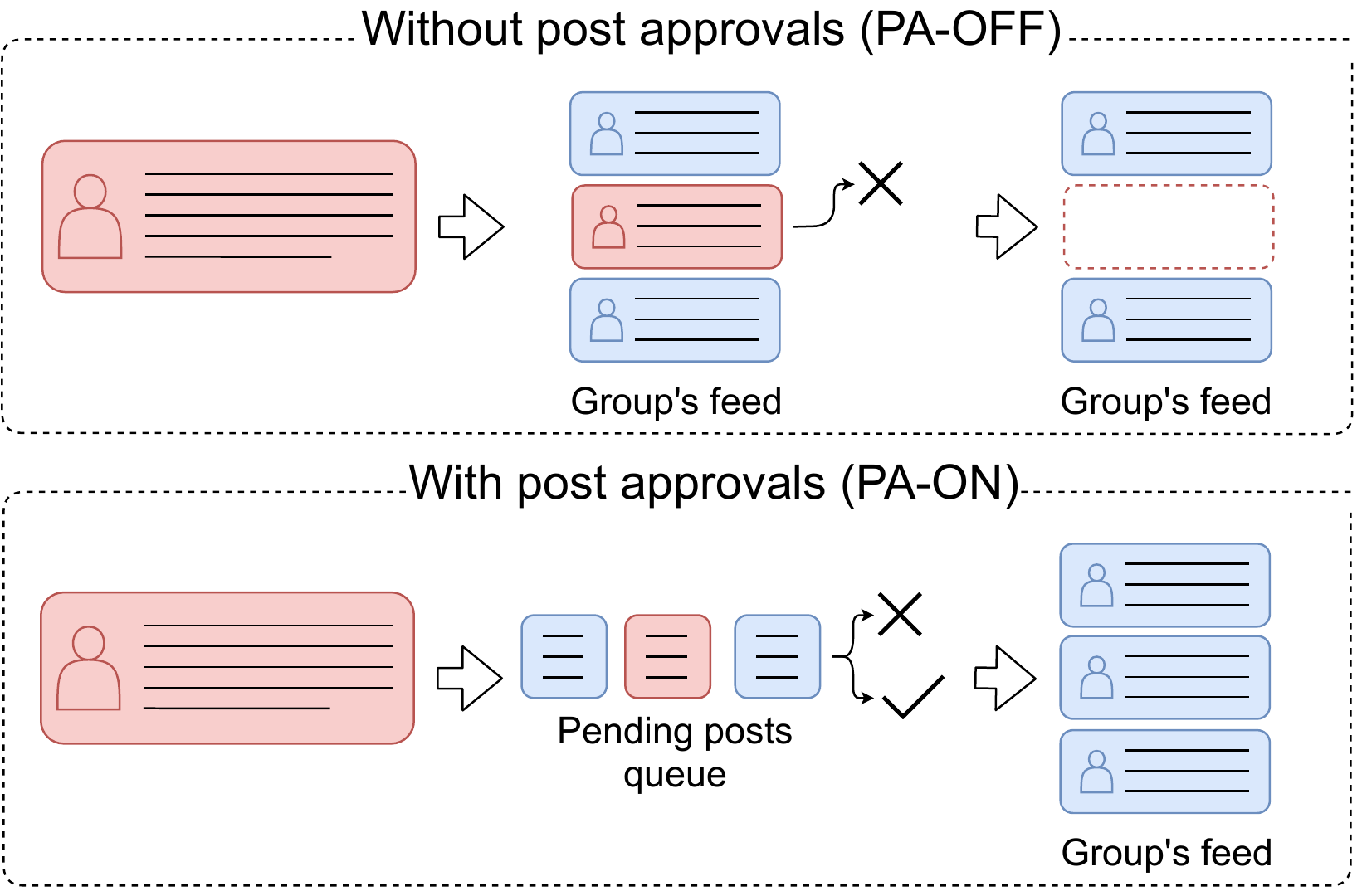}
    \caption{Post approvals allow posts that violate a community's guidelines (in red) to be filtered \emph{before} other members in the community see them.
    Without post approvals, these posts can only be moderated \emph{after} they are posted in the group.
    }
    \vspace{2mm}
    \label{fig:dic}
\end{figure}

Online communities are partially shaped by the design affordances of the platforms they inhabit~\cite{bucher2018affordances}.
In Facebook Groups, administrators can turn on ``post approvals,'' a setting that requires members' posts to be accepted by community leaders (i.e., administrators and moderators) before others in the group can see and interact with them~\cite{pa22}.
This setting changes \emph{when} norms are enforced in a community or group, as illustrated in Figure~\ref{fig:dic}.
If the setting is turned off, community leaders must reactively moderate the posts in the community, e.g., by browsing posts in the group as they appear or by responding to reports from other members or the platform.
If the setting is turned on, leaders can proactively moderate the community, prescreening posts that are low quality or that break the rules.

Well-moderated spaces are more attractive to users~\cite{wise2006moderation} and can improve the quality of users' contributions~\cite{cosley2005oversight}.
However, over-enforcement of rules can discourage participation~\cite{jhaver2019does, kiene2016surviving}, and moderation creates more work for leaders~\cite{lo2018all, dosono2019moderation}.
Thus, post approvals, a proactive moderation strategy, involves several trade-offs.
On the one hand, it may prevent harm caused by violations of a community's guidelines and improve members' overall experience.
On the other hand, it introduces participation friction and may increase leaders' workloads.

\xhdr{Present work} 
This paper presents an observational study of the adoption and the impact of post approvals in online communities.
We ask:

\begin{itemize}
    \item \textbf{RQ1} What leads communities to adopt post approvals?
    \item \textbf{RQ2} How do post approvals shape user activity and moderation in online communities?
    \item \textbf{RQ3} Does the impact of post approvals depend on community properties and on how the setting is used?
\end{itemize}

Using a longitudinal dataset of user activity\hyp{} and moderation\hyp{}related traces from 233,402 Facebook Groups from March to July 2021, we compared communities that enabled post approvals (PA-ON; $n=8{,}767$) to communities that did not change any moderation-related settings (PA-OFF; $n=224{,}635$).

To examine the factors that led to the adoption of post approvals \textbf{(RQ1)}, we studied activity in PA-ON and PA-OFF communities in the 4 weeks before the former enabled the setting.
During this period, PA-ON communities experienced greater growth in user activity (\eg, \emph{Comments}) and moderation (\eg, \emph{Posts reported}) compared to PA-OFF communities.
Further, right before PA-ON communities enabled post approvals, they experienced a sudden increase in moderation, which may have been the final straw that led administrators to turn on the setting.
These findings continue to hold when using propensity score matching to control for initial baseline user and moderation activity, 
and are further confirmed when examining how user activity or moderation predicts if post approvals will be turned on in future weeks.

To study how post approvals shape online communities \textbf{(RQ2)}, we matched PA-ON and PA-OFF communities on user activity and moderation traces in the 4 weeks prior to post approvals being turned on and compared differences in their subsequent activity.
We found that, while fewer posts were shared in groups that enabled post approvals, the posts that were shared received more comments, more reactions, more time spent, and fewer reports,
suggesting improvements in the quality of content being posted.
Further, post approvals did not significantly increase the average time leaders spent in their groups, though groups that enabled the setting tended to increase their moderation team.

Last, post approvals may differently impact a community depending on its properties and on how post approvals are used in practice \textbf{(RQ3)}.
To understand these differences, we studied how the effects of post approvals varied with group size (\ie, how many members there were in a group), leaders' response time for submitted posts (\ie, how much time did it take for a post to be approved) and the post approval rate (\ie, what fraction of posts submitted in a given group were approved).

For all three factors, we found significant interactions with time spent by leaders in the group and with changes in activity in the group following the adoption of the setting.
Leaders spent significantly more time after adopting post approvals in larger groups, groups with higher post approval rates, and groups with faster response times.
There were sharper decreases in the number of posts and increases in the number of comments, reactions, and time spent per post in larger groups, groups with lower post approval rates, and groups with slower response times.
Still, other changes persisted across different communities.
Independent of group size, response time, or approval rate, the fraction of posts reported decreased significantly after post approvals was adopted. This suggests that regardless of how post approvals was enforced, the setting nonetheless reduced content perceived by members as problematic or rule-breaking.

Overall, our findings suggest that post approvals substantially change how online communities work and that the setting creates communities centered around fewer, higher-quality posts.
These insights may guide improvements to community-level moderation processes
and the quasi-experimental approach we adopted can be easily extended to analyze other opt-in features provided by social media platforms.

\section{Related Work}

Moderation in online communities increases their attractiveness to newcomers~\cite{wise2006moderation}, improves the quality of contributions~\cite{cosley2005oversight}, and decreases anti-social behavior~\cite{seering2017shaping}.

Nonetheless, effective moderation is difficult -- platforms experience several challenges related to the scale, the legitimacy, and the contextual nature of content moderation~\cite{gillespie2018custodians, filgueiras2021digital}.
Beyond work to better predict when content may violate community guidelines~\cite{schmidt2019survey, papadamou2020disturbed}, community-oriented social media (and moderation) has been suggested as part of the solution to these challenges because community leaders may better incorporate local and cultural context into moderation decisions~\cite{seering2020reconsidering} and because the decisions taken would be considered more legitimate~\cite{filgueiras2021digital}.
To this end, some research has examined how moderators engage and regulate their communities~\cite{seering2019moderator} by developing specific design guidance from fundamental theories in the social sciences~\cite{kraut2012building}.

Most relevant to the present work are existing studies that explored how technological affordances provided by platforms shape content moderation. For example, participation controls~\cite{kraut2012building} limit what specific users are allowed to see or do within a social media platform or a specific online community.
In the development of open-source software, collaborators receive ``commit rights'' as they offer evidence of their technical expertise~\cite{ducheneaut2005socialization};
on PalTalk (an early video group chat service), moderators could impose ``activity quotas'' to chat room users, limiting their participation~\cite{kraut2012building};
on Twitch, moderation includes chat ``modes'' that change how users can participate, for instance allowing only emotes to be sent~\cite{seering2017shaping};
on Reddit, \citet{jhaver2019human} studied the usage of \textit{AutoModerator}, a system that allows moderators to define ``rules'' to be automatically applied to posts in their communities.

This work examines post approvals, a participation control that is central to community-level moderation in Facebook Groups but whose specific effects have not yet been systematically studied.
Post approvals change the dynamics of content moderation by allowing community leaders to \textit{proactively} moderate posts before they ever land in the communities' feeds.
Moreover, when someone attempts to contribute to a community with post approvals turned on, posts may take hours or even days to get published (if they do).
Communities could thrive in the better-moderated spaces enabled by post approvals~\cite{wise2006moderation} and the participation friction could disrupt mindless interactions~\cite{mejtoft2019design}.
However, the setting could also discourage participation~\cite{kiene2016surviving} and create unnecessary work for leaders~\cite{lo2018all,dosono2019moderation}.

Studying the impact of post approvals (and other participation controls) in online communities can help create better governance practices and further our understanding of how participation friction and proactive moderation can improve online spaces. 

\section{Data}
\label{sec:data}
Between March 28, 2021, and July 11, 2021, we collected data on
1)~communities that turned on post approvals and did not change other moderation-related settings (PA-ON; $n=8{,}767$); and 
2)~a random sample (50\%) of communities that did not change \emph{any} moderation-related setting (PA-OFF; $n=224{,}635$).
For PA-ON groups, we considered only communities that enabled post approvals at least 28 days after the start of the study period and at least 28 days before its end.
For both PA-ON and PA-OFF groups, we considered only communities with 128 or more members and at least one comment and one post over any 7-day window. 

We analyzed PA-ON communities relative to when they turned on post approvals, referring to the day when they enabled the setting as day 0.
For PA-OFF communities, we randomly assigned a pseudo-intervention date drawn from the distribution of dates (day and hour) when PA-ON groups enabled post approvals (\cf \Appref{app:adplots}; \Figref{fig:day_hour}).
We considered the set of variables described in \Tabref{tab:cov} in the 28 days before and after each intervention, for a total of 57 days (from $-28$ to 28). 
Some variables are 
1)~time-dependent, capturing group activity and group moderation (\eg, number of posts, number of posts deleted),
while others are 2)~time-invariant, capturing group topic, demographics, and moderation settings
(\eg, group visibility, group category, if a group was a buy-and-sell group, \etc). 

All data was de-identified and analyzed in aggregate, and no individual-level data was viewed by the researchers. 
In the analyses that follow, variables were 95\%-winsorized (\ie, the 2.5\% smallest and largest values were replaced with the most extreme remaining values~\cite{wilcox2011introduction}) prior to aggregation unless otherwise stated. This ensured that trends/effects were not dominated by a few large groups. Nonetheless, results were qualitatively similar without winsorization.

\section{What Leads to the Adoption of Post Approvals?}
\label{sec:rq1}

This section examines \textit{why} communities adopt post approvals to begin with (\textbf{RQ1}).
We focus on what happens \textit{before} the setting was enabled, contrasting PA-ON and PA-OFF groups.
All analyses in this section were done at the group level, with each group weighted equally.

\begin{table}[H]
    \centering
    \footnotesize
    \begin{tabular}{p{2cm}p{5.5cm}}
    \toprule
    \multicolumn{2}{c}{\textbf{Group characteristics} (not time-dependent) }\\ \midrule
    \textbf{Visibility$^\star$} &  Whether group is private or public.\\   \midrule
    \textbf{Join approvals$^\star$} & Whether leaders have to manually approve new members.  \\ \midrule
    \textbf{Average age$^\star$} & Average age of the members in the group.  \\ \midrule
    \textbf{\% women$^\star$} & The percentage of women in the group.  \\ \midrule
    \textbf{Buy-\&-Sell$^\star$} & Whether the group is a buy and sell group or not (specified by admin).   \\ \midrule
    \textbf{Group categories$^\star$} & Lexical categories obtained from the groups' description and title using Empath~\cite{fast2016empath}. See \Appref{app:topics} for details.  \\ \midrule
    \multicolumn{2}{c}{\textbf{Moderation-related}} \\ \midrule
    \textbf{Moderating TS} & Average time leaders spent in moderation-related interfaces (e.g., approving posts).\\ \midrule
    \textbf{Leader TS} & Average time leaders spent in the group.\\   \midrule
    \textbf{Members Removed} & Number of members removed. \\ \midrule
    \textbf{Posts deleted} & Number of posts by regular members deleted by leaders. \\ \midrule
    \textbf{Posts reported} & Number of posts reported in the community by users. \\ \midrule
    \textbf{Num leaders} & Number of leaders in the community. \\  \midrule
    \multicolumn{2}{c}{\textbf{Activity-related}} \\ \midrule
    \textbf{Posts} & Number of posts. \\  \midrule
    \textbf{Comments} & Number of comments. \\  \midrule
    \textbf{Time spent} & Total time users spent browsing posts in the group (in hours).
    \\  \midrule
    \textbf{Reactions} & Number of Likes and of other reactions (Sad, Happy, Wow, Laugh, Angry). \\ \midrule
    \textbf{Num members} & Number of members in the community. \\  \bottomrule
    \end{tabular}
    \caption{Description of the group-level variables considered in this paper. Variables marked with a star ($\star$) were measured on the day prior to the intervention (for PA-ON groups) or the pseudo-intervention (for PA-OFF groups). They were not analyzed in the result sections of this paper, but were used in the matching to ensure the two sets of communities were comparable (\cf, \Appref{app:psm}).}
    \label{tab:cov}
\end{table}

\begin{figure*}[t]
    \centering
    
    \begin{minipage}[t]{0.49\linewidth}
    \subcaption{All} 
    \includegraphics[width=\linewidth]{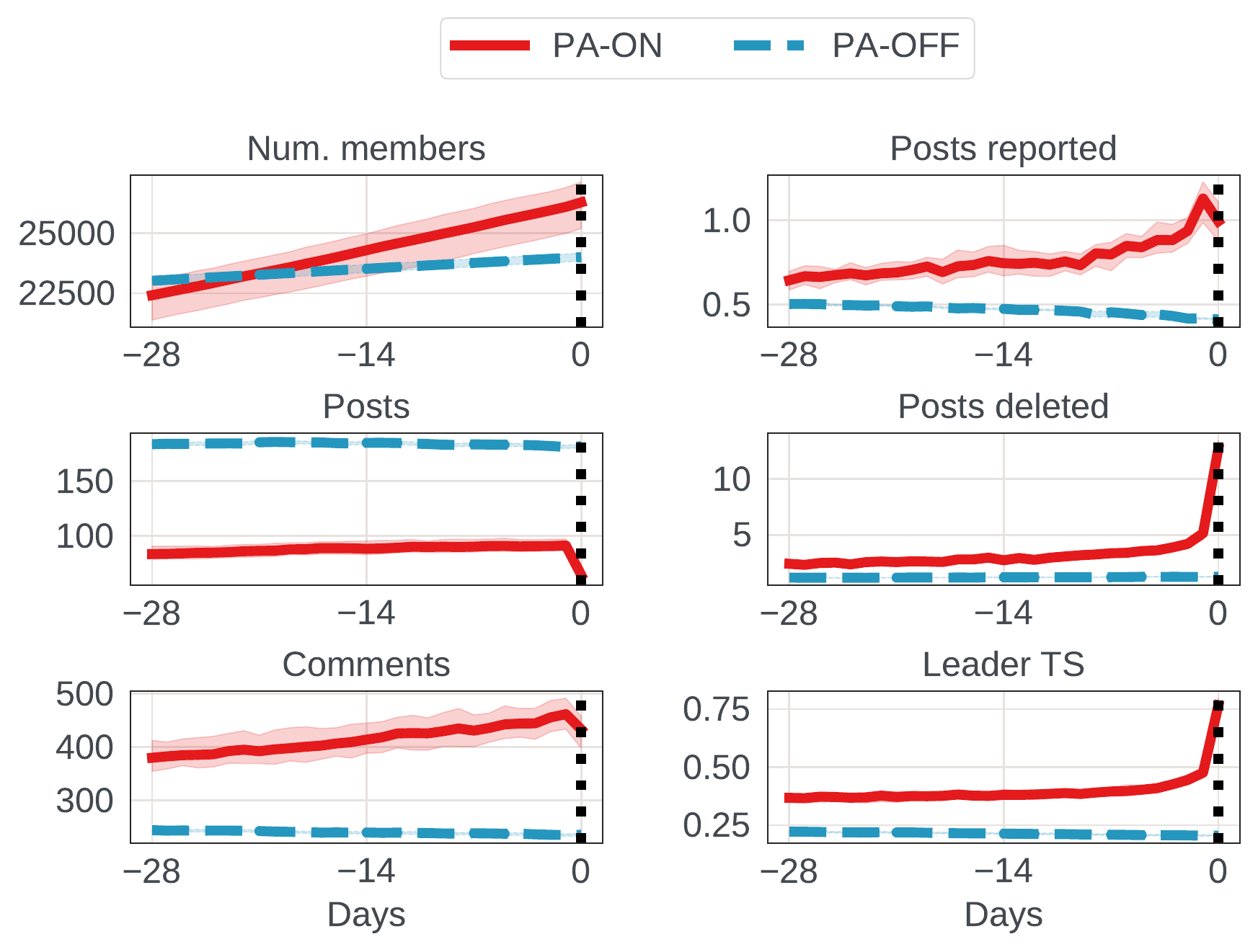}
    \end{minipage}%
    \hspace{1mm}
    \begin{minipage}[t]{0.49\linewidth}
    \subcaption{Matched} 
    \includegraphics[width=\linewidth]{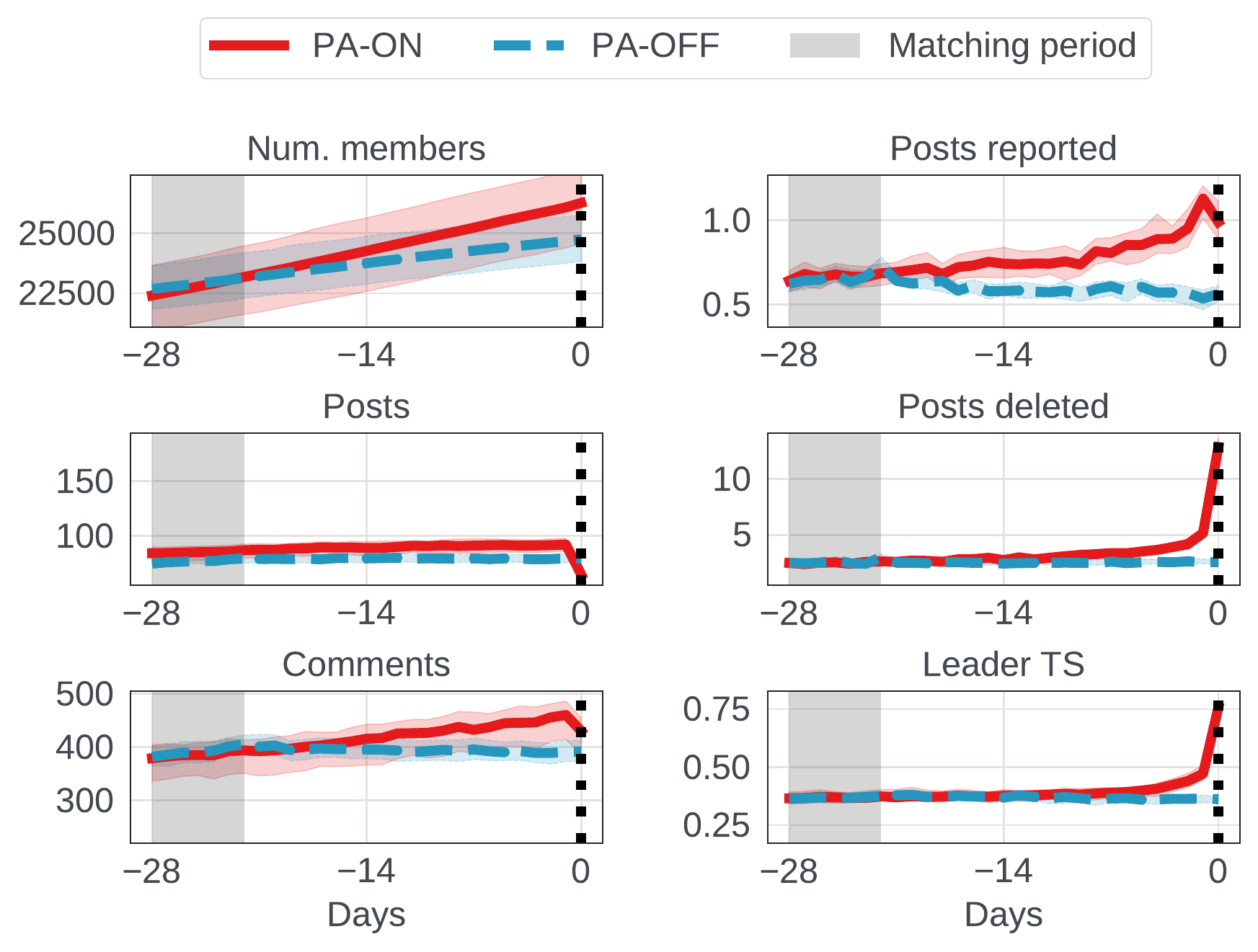}
    \end{minipage}
    \caption{Average values for user activity\hyp{} and moderation\hyp{}related variables in the four weeks before communities enabled post approvals. 
    Values for communities that enabled post approvals (PA-ON) are in red and those for communities that did not (PA-OFF) are in blue.
    For PA-OFF communities, day 0 corresponds to a pseudo-intervention date selected at random.
    We show trends for all communities in our dataset in \textit{(a)} (PA-ON $n=8{,}767$; PA-OFF $n=224{,}635$) and for matched pairs of communities in \textit{(b)} (PA-ON/PA-OFF $n=8{,}643$).
    The period when matching was done is marked in gray. Error bars represent 95\% CIs.
    }
    \label{fig:case_control}    
\end{figure*}

\xhdr{Case--control analysis}
Our first analysis follows a case--control design~\cite{schlesselman1982case}: we compared user activity and moderation traces of groups that enabled post approvals (PA-ON; the ``case'') with those that did not change any moderation settings (PA-OFF; the ``control'').
We considered three variables related to user activity (\emph{Num Members}, \emph{Posts}, and \emph{Comments}) and three variables related to moderation activity (\emph{Posts reported}, \emph{Posts deleted}, \emph{Leader TS}) in the 28 days before the intervention (\cf \Tabref{tab:cov} for descriptions).
We refer to this scenario as ``all'' since, in what follows, we examine a subset of this data corresponding to matched pairs of PA-ON and PA-OFF groups.

\Figref{fig:case_control}a shows the average value of each of the variables mentioned above.
We found significant differences between PA-ON and PA-OFF groups that are consistent across the 28-day period considered ($p<10^{-4}$ for independent t-tests conducted each day).
PA-ON groups have significantly more reported and deleted posts, more comments, and fewer posts than PA-OFF groups.
Leaders also spent more time in PA-ON groups than in PA-OFF groups.

Temporal trends also differed significantly between the two sets of groups: PA-ON groups experienced larger increases in all considered variables in the weeks before enabling post approvals.
For moderation-related metrics, we observed a sharp spike on the day (or in the case of posts reported, on the day before) post approvals was turned on. 
These changes were not observed in PA-OFF groups.

The shifts in user activity (\eg, \emph{Comments}) and in moderation (\eg, \emph{Posts deleted}) before post approvals was turned on suggests that leaders enable the setting in response to new (and perhaps more chaotic) group dynamics.
Specifically, the setting was commonly enabled in groups that were quickly growing and that experienced a surge in moderation-related events, which may have been the final straw that led administrators to enable post approvals.
This finding is consistent with prior work suggesting that major changes in moderation (\eg, changing settings, creating new rules) happen in reaction to problems that emerge \cite{seering2019moderator}.

\xhdr{Matched analysis}
While indicative, the previous analysis conflates two factors. 
Not only do PA-ON and PA-OFF communities differ in their baseline user and moderation activity, but they also differ in the way the studied variables change over time. 
Thus, observed differences in temporal trends may come from the fact that groups that adopt post approvals are different from those that do not.
To more fairly compare these communities, we matched PA-ON and PA-OFF groups on user activity and moderation-related metrics between days $-28$ and $-22$.
This matching ensures that communities were similar in the first week of the study period.
Specifically, we performed one-to-one propensity score matching of PA-ON and PA-OFF communities using moderation and activity-related variables, as well as general group characteristics (\eg, \emph{Group categories}). Details of the matching procedure can be found in \Appref{app:psm}.

After performing this matching, we repeated the same analysis as in the previous subsection (\Figref{fig:case_control}b).
Again, we found that PA-ON groups experienced a gradual increase in moderation-related traces which was accentuated right before post approvals was turned on — changes that were not observed in PA-OFF groups.
Differences in user activity were subtler. 
Both PA-ON and PA-OFF groups experienced growth in the number of members, but this growth was higher for PA-ON groups.
Moreover, PA-ON groups experienced a significant increase in the number of daily comments received, while comments received remained largely unchanged in PA-OFF groups.
Last, in both PA-ON and PA-OFF groups the number of posts increased slightly during the 28 days considered.

Overall, the matched case--control analysis confirms that there are differences in the temporal trends of moderation and user activity of PA-ON and PA-OFF communities. 
Even when considering communities that were initially similar, PA-ON communities experienced larger increases in user and moderation activity prior to the day when they turned on post approvals.

\begin{figure}[t]
    \centering
    \includegraphics[width=\linewidth]{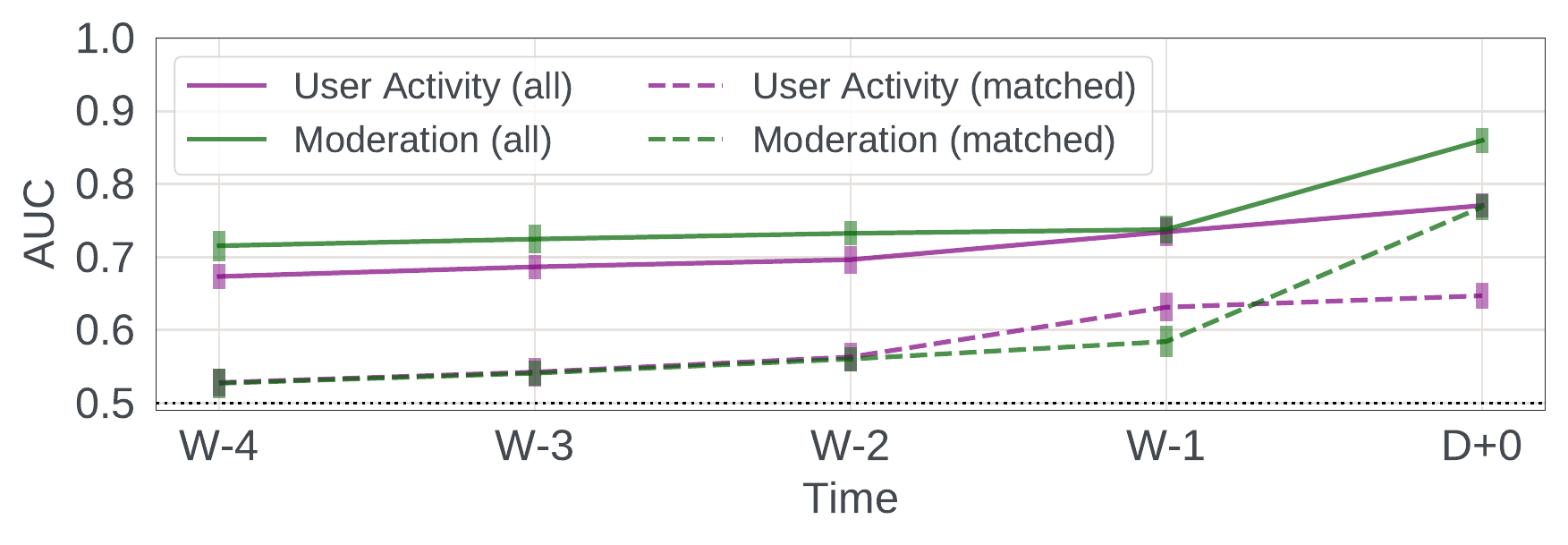}
    \caption{AUCs for classifiers trained to distinguish PA-ON and PA-OFF groups using data from different time spans and different sets of features.
    Error bars represent 95\% CIs obtained through a 20-fold cross validation.
    }
    \label{fig:class}
\end{figure}

\xhdr{Predicting if post approvals will be turned on in future weeks}
While findings thus far indicate that both changes in user activity and in moderation precede the use of post approvals, is one a stronger indicator than the other?
And how far in advance might they predict the adoption of the setting? 
To answer these questions, we examined if user activity or moderation can be used to distinguish PA-ON groups from PA-OFF groups.

We created two balanced samples of groups.
The first sample (\emph{all}) comprised 10,000 groups — half PA-ON and half PA-OFF. 
The second (\textit{matched}) comprised the same PA-ON groups as in the \emph{all} sample, while corresponding PA-OFF groups were obtained using one-to-one propensity score matching previously described.
We considered two sets of features (group activity-related and moderation-related; \cf \Tabref{tab:cov}) and five time spans,%
\footnote{
Days -28 to -22 (week -4; W-4),
-21 to -15 (week -3; W-3),
-14 to -8 (week -2; W-2), 
-7 to -1 (week -1; W-1), and
day 0 (D+0).}
calculating the value of each feature in each time span by taking its average.

We trained Gradient Boosting classifiers to distinguish PA-ON and PA-OFF groups, varying the feature set and the time span used.
\Figref{fig:class} shows the AUC of the classifiers trained in each of the different settings (all \vs matched; moderation \vs activity) 
considering the features up to the time specified on the $x$-axis. For instance, the points shown above $x=$ W-3 correspond to the AUC of classifiers trained with features associated with W-4 and W-3. 

For the \emph{all} sample, we found that moderation features were more predictive in earlier weeks (W-4 to W-2). However, for $x=$ W-1, there was an increase in the AUC of the classifier trained with activity features.
We observed a similar pattern in the \emph{matched} sample: classifiers started with similar AUC values at $x=$ W-4 (user activity: 0.53 AUC \vs moderation: 0.53 AUC), but the classifiers trained with activity features saw a larger increase in performance at $x=$ W-1 (0.58 for activity \vs  0.63 for moderation).
Including data up until the time of intervention, $x=$ D+0, the performance of classifiers trained with moderation features increased sharply (\eg, in the \emph{matched} sample: 0.65 activity \vs  0.77 moderation).
Overall, these results suggest that the adoption of post approvals was associated with gradual changes in user activity in the weeks before the adoption of the setting and sudden changes in moderation activity on the day the setting was enabled.
Repeating this analysis but instead training classifiers using features belonging to each individual time span (e.g., using only W-1 vs. using W-4 to W-1 for prediction) resulted in qualitatively similar findings.

\section{How do Post Approvals Shape Online Communities?}
\label{sec:rq2}

Having explored changes in user activity\hyp{} and moderation-related signals that \emph{precede} the adoption of post approvals, we now turn our attention to what happens \emph{after} communities choose to adopt the setting.
Here, we examine how user and moderation activity in online communities change following the adoption of the setting (\textbf{RQ2)}. To do so, we matched communities that turned on post approvals (PA-ON) with similar communities that did not (PA-OFF) and validated the observed differences using regression.
To obtain this matching, we performed one-to-one propensity score matching on moderation and activity\hyp{}related variables as well as general group characteristics. Matching was done across the entire pre-intervention period (day $-28$ up to day 0 right before the intervention). See Appendix A for details.
All analyses in this section were done at the group-level, with each group weighted equally.

\begin{figure}[t]
    \centering
    \includegraphics[width=\linewidth]{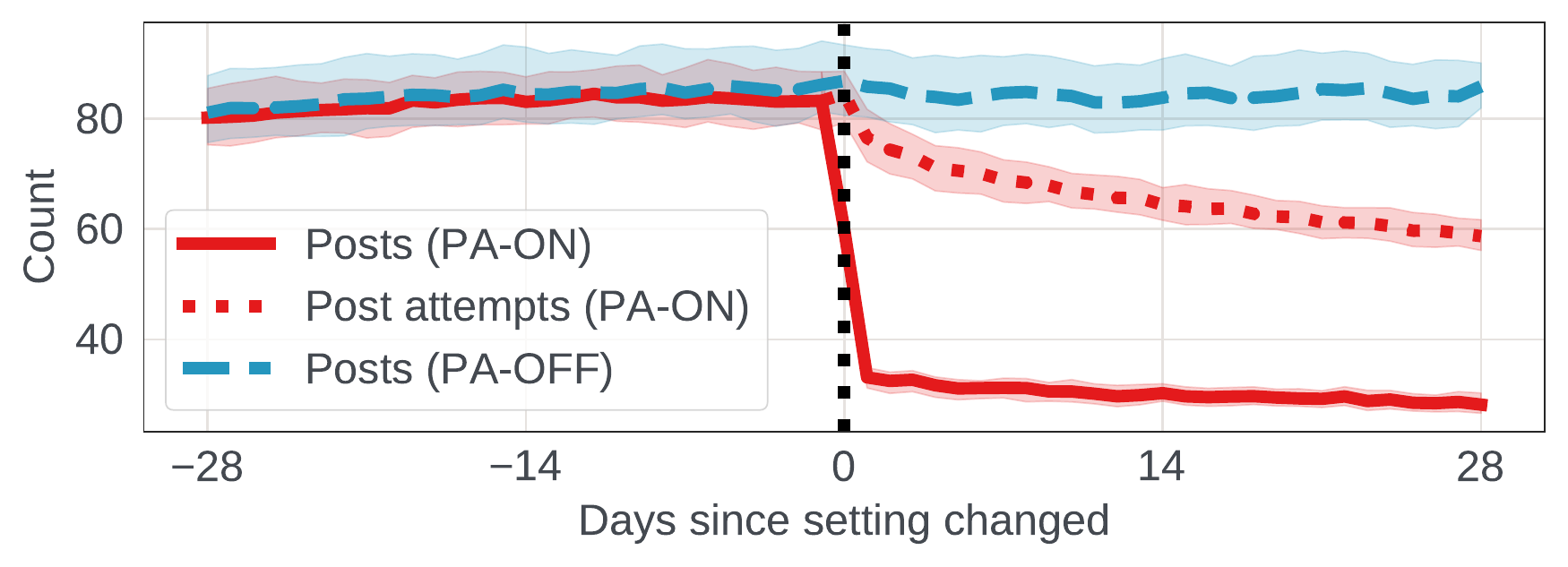}
    \caption{The average number of posts in PA-ON (solid red) and PA-OFF (dashed blue) communities.
    For PA-ON communities, the average number of post attempts after post approvals was enabled is shown in dotted red. Error bars represent bootstrapped 95\% CIs.
}
\label{fig:posts1}
\end{figure}

\xhdr{Posts}
First, we examined how posting behavior changed after the adoption of post approvals.
In \Figref{fig:posts1}, we show both the number of posts that were actually published (\emph{Posts}), but also, for PA-ON communities, the number of post attempts, i.e., post requests initiated by regular group members following the adoption of post approvals.
For PA-ON communities, we found a significant decrease in the number of posts following the adoption of post approvals.
The average number of posts went from roughly 80 posts a day pre\hyp{}intervention to around 30 posts a day post\hyp{}intervention, a decrease that was not observed in the matched set of PA-OFF communities.

What explains this decrease in posting? Was it because posts were being filtered? Or were people more hesitant to even post?
To understand the relative contribution of these factors, we examined two corresponding quantities that make up the decrease in posting:
1)~the difference between the number of posts in the control (PA-OFF) and treatment setting (PA-ON) (blue line vs. dotted red line); and 
2) the difference between the number of post attempts and actual posts in PA-ON communities (dotted red line vs. solid red line).
We observed a gradual decrease in the average number of posts submitted (the first component mentioned above), from 76 posts a day on day 1 to 58 posts a day on day 28.
However, the fraction of posts approved per community (the second component) remained largely stable at around 63\% of posts (note that this was calculated without winsorization and per group, instead of dividing the overall averages; \cf \Figref{fig:pct}, \Appref{app:adplots}).
These findings suggest that post approvals reduce the number of posts by directly filtering out undesired posts, but also by reducing the likelihood of people to attempt to post in the first place.

\begin{figure}
    \centering
    \includegraphics[width=\linewidth]{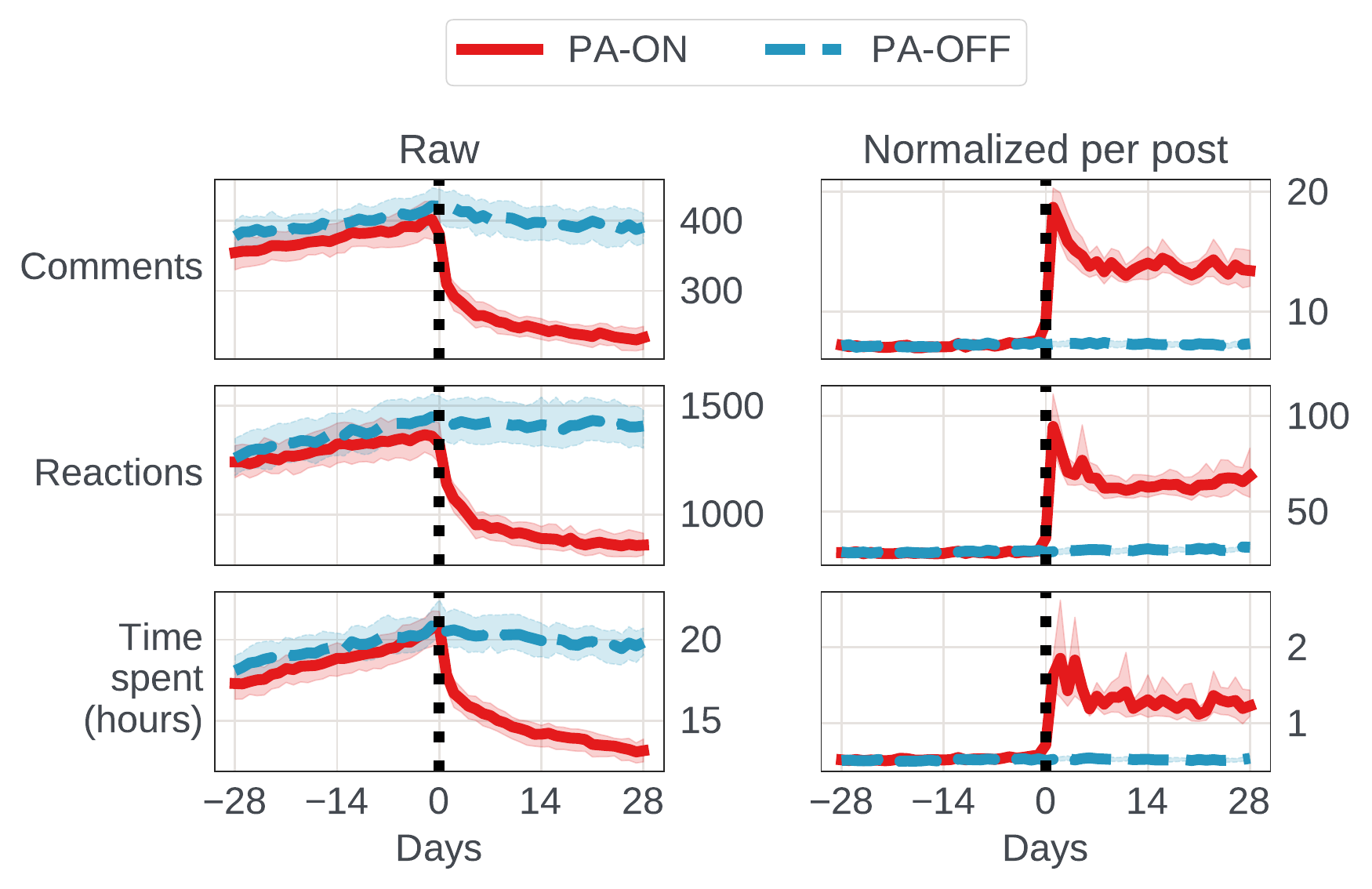}
    \caption{User activity-related signals before and after the adoption of the post approval setting. Signals are shown both in absolute terms (left) and normalized per number of posts (right). Error bars represent bootstrapped 95\% CIs.}
    \label{fig:posts2}
\end{figure}

\xhdr{Other user activity-related signals} 
Second, we looked at other user activity-related metrics (\emph{Comments}, \emph{Reactions}, and \emph{Time spent}). These are shown in \Figref{fig:posts2} both in absolute terms (first column) and normalized per number of posts (second column). 
Comparing PA-ON with PA-OFF communities, there was an absolute \emph{decrease} but relative \emph{increase} in all of these user activity metrics for PA-ON communities following the enabling of post approvals.
In other words, there were fewer posts, but each post received, on average, more comments, reactions, and time spent.

For instance, before the intervention (day $-1$), PA-ON and PA-OFF communities received an average of around 402 and 421 daily comments and around 7.6 and 7.4 comments per post. 
(Note that although the two sets of communities are matched, their averages are not perfectly identical.)
After day 0, when PA-ON communities enabled post approvals, the number of daily comments declined substantially, reaching an average of 233 daily comments on day 28.
Meanwhile, the number of comments per post nearly doubled to around 13.4.
This change was not observed in the matched PA-OFF groups.

\begin{figure}
    \centering
    \includegraphics[width=\linewidth]{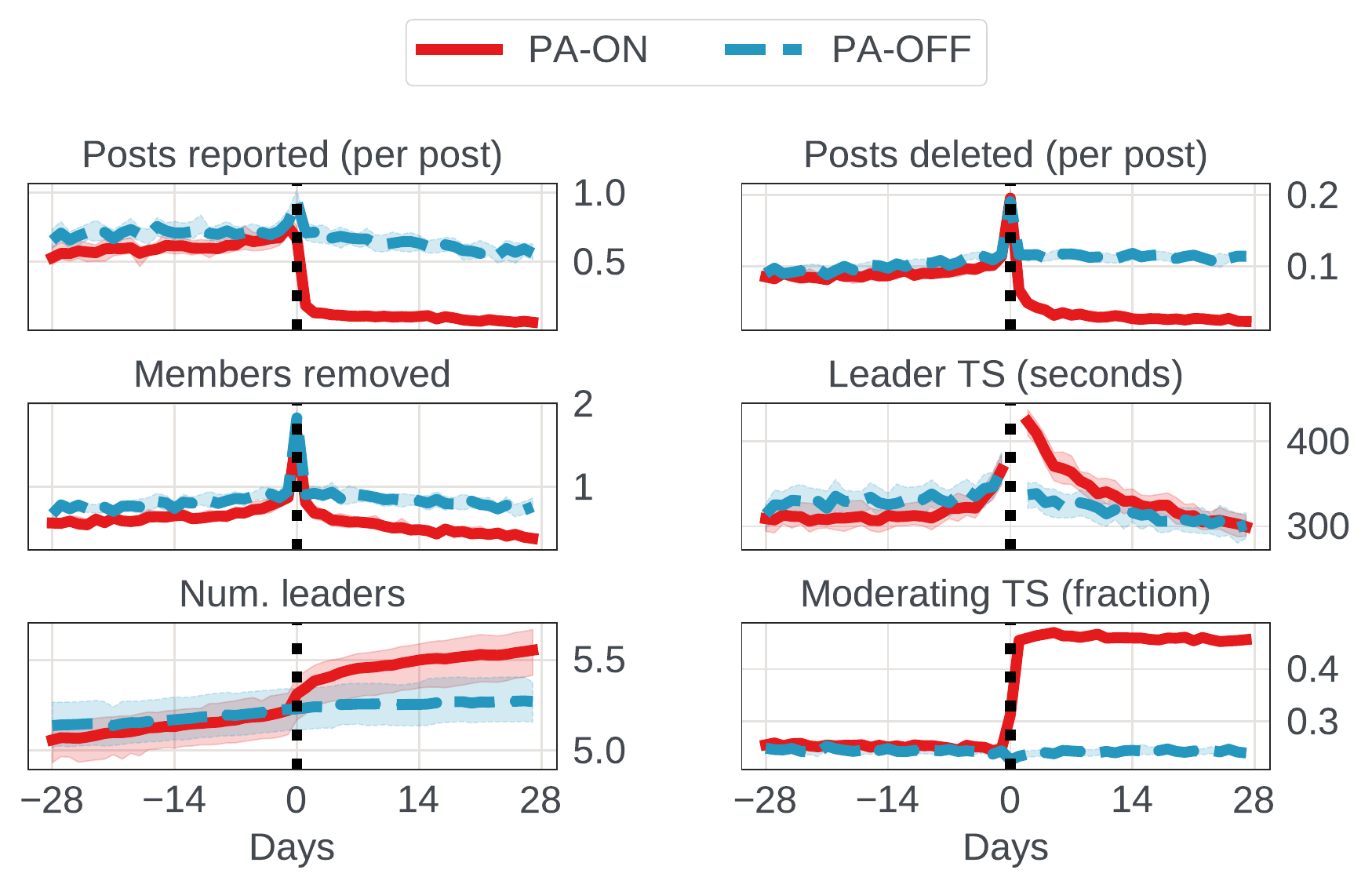}
    \caption{Moderation-related signals before and after the adoption of post approvals. Error bars represent 95\% CI. We omit days 0 and 1 from the plot showing the \emph{Leader TS}, as it contains a sharp peak.}
    \label{fig:posts3}
\end{figure}

\xhdr{Moderation-related signals} 
Third, in \Figref{fig:posts3}, we examined moderation-related metrics -- \emph{Posts reported}, \emph{Posts deleted}, \emph{Leader TS} and \emph{Moderating TS} (\cf \Tabref{tab:cov} for descriptions).
We normalized the number of posts reported and deleted by the total number of posts, and moderating time spent by  the total leader time spent.
Recall that the moderating time spent encompasses activities such as responding to reported content and, notably, approving posts (if the post approvals setting is turned on).

Following day 0, PA-ON groups had fewer members removed and fewer posts reported/deleted (per post) than PA-OFF communities. For example, the percentage of posts reported decreased from around 0.75\% pre-intervention to 0.10\% post-intervention for PA-ON groups. This decrease was larger than that for matched PA-OFF groups (from 0.78\% to 0.60\%).
Around the time that post approvals was enabled for PA-ON groups, time spent per admin increased substantially, likely because leaders were getting used to the new moderation style.
However, by week 4, time spent per leader in PA-ON groups had returned to pre-intervention levels, though PA-ON groups had also tended to appoint new leaders.
Examining the fraction of time spent in admin surfaces, leaders went from spending around 25\% of their time using group moderation tools to around 45\%, suggesting that leaders spent a substantial fraction of their time approving posts.
While this increase appears large, group leaders in groups without post approvals may nonetheless be informally vetting posts by browsing posts in a group as they appear.

\begin{figure}
    \centering
    \includegraphics[width=\linewidth]{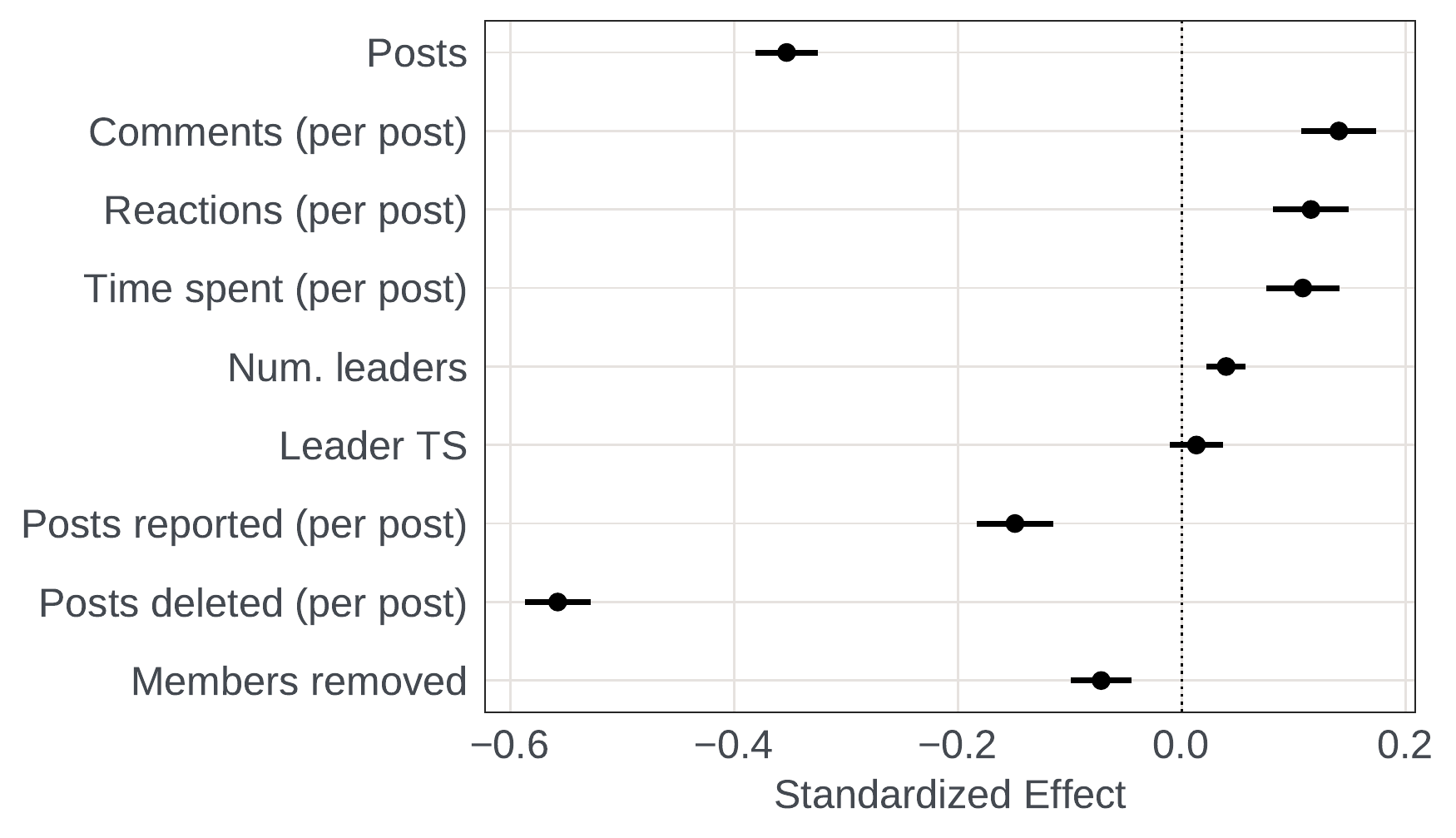}
    \caption{Standardized effect of enabling post approvals on user activity\hyp{} and moderation\hyp{}related variables. Error bars represent 95\% confidence intervals. Data was not winsorized prior to this analysis.}
    \label{fig:posts4}
\end{figure}

\begin{figure*}
    \centering
    \includegraphics[width=\linewidth]{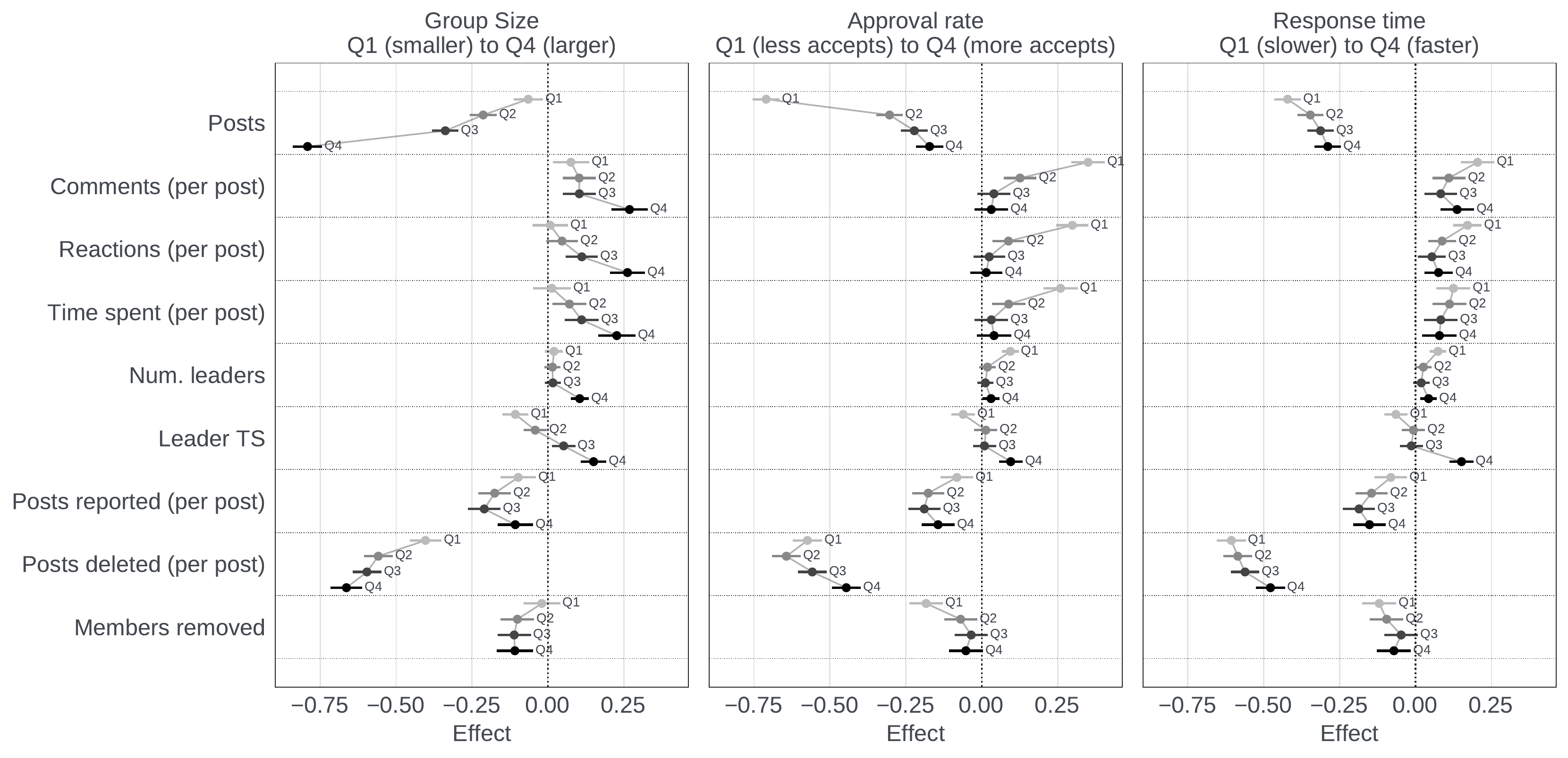}
    \caption{Effects of enabling post approvals for different stratifications of the data. Here, we report quartile-specific effects for group size (\ie, number of members in a group; first column), the post approval rate (\ie, percentage of posts that get approved in a group; second column) and the response time (\ie, average time taken to accept posts in a group; third column). Data was not winsorized prior to this analysis. Error bars represent 95\% CIs.}
    \label{fig:posts5}
\end{figure*}

\xhdr{Regression analysis}
Previously, we compared the user and moderation signals between PA-ON and PA-OFF communities before and after the post approvals are turned on after matching. 
Here, we performed a more rigorous analysis of the same signals under a regression framework.

We considered the average value of each variable of interest in week $-4$ (days $-28$ to $-22$) and week 4 (days 22 to 28). Then, for the variables in the post-intervention period, we estimate the impact of adopting post approvals using a linear model:
\begin{equation}
\label{eq:reg}
    y = \hm{\alpha} \mathbf{X} + \beta \, \hm{1}_{\text{[PA-ON = True]}},
\end{equation}
where $y$ represents the average value of one of the variables we studied in week 4 (\ie, after the intervention), \eg post approvals, $\mathbf{X}$ represents an array with all the variables we did the matching with in week $-4$ (\ie, before the intervention), and $\beta$ represents the coefficient associated with turning on post approvals, as it multiplies an indicator variable that equals 1 for PA-ON, and 0 for PA-OFF, communities.

To facilitate interpretability, we standardized the dependent variable $y$, so that the coefficients represent (pooled) standard deviations.
The coefficient $\beta$ captures the difference between our treatment and control groups in the matched setting.
Since the coefficient is associated with an indicator variable, the effects reported represent the differences between PA-ON and PA-OFF groups in standard deviations.
We report $\beta$ for all outcomes of interest in \Figref{fig:posts4}. 

This analysis largely confirms results shown in \Figref{fig:posts1}, \Figref{fig:posts2}, and \Figref{fig:posts3}.
The use of post approvals was significantly associated with a reduction in the number of posts ($-0.35$ standard deviations) but an increase in the number of comments, reactions, and time spent per post (e.g., the number of comments per post increased by 0.14 SDs).
Use of the setting was also associated with a decrease in the number of posts reported per post ($-0.15$ SDs), posts deleted per post ($-0.55$ SDs) and number of members removed ($-0.07$ SDs).
Taken along with the previous analyses, these results suggest that the setting improves the quality of posts.
Further, post approvals do not significantly increase the average time leaders spend in the group, although groups that enable the setting tend to increase the size of their leadership team (around 0.04 SDs).

\section{Heterogeneity of Post Approvals} 
\label{sec:het}
While post approvals change how online communities function, the effect of the setting may vary by group size as well as how it is used.
For example, we found that, on average, community leaders do not spend more time in their communities following the adoption of the setting. 
Yet, this may not be the case for all groups: very large groups (with possibly hundreds of daily post attempts) may actually require more time from leaders after the setting is turned on, while smaller groups may require less time.
Thus, we analyzed the impact of post approvals in communities with 1)~different member counts, as well as in communities that 2)~approved different fractions of the posts submitted (approval rate); and 3)~took a different amount of time to approve posts (response time).

For each of the aforementioned variables (number of members; response time and approval rate), we divided PA-ON communities into 4 quartiles.%
\footnote{Defined by three points:
for approval rate: 0.45, 0.69, 0.84; 
for response time: 1.2, 2.7, 5.75 (hours);
for group size: 2850, 8500, 24000 (rounded) Approval rate and response time were measured across the entire post-intervention period. Group size was measured on day 0.}
Then, we used the same regression setup depicted in Equation \eqref{eq:reg}, but estimated the effect of post approvals separately for communities in each of the quartiles. This amounts to a linear model of the form:
\begin{equation}
\label{eq:reg2}
    y = \hm{\alpha} \mathbf{X} + \sum_{i=1}^{4} \beta_i \, \hm{1}_{\text{[PA-ON = True and Quartile = $i$]}},
\end{equation}
where $\beta_i$ is the effect for groups in a given quartile. We ran a different regression for each of the three setups described above (group size, approval rate, and response time).

All three factors had noteworthy interactions with user activity\hyp{} and moderation-related signals (\cf\ \Figref{fig:posts5}).

\xhdr{Time spent by leaders} 
Leaders in larger groups (group size Q4), groups with lower approval rates (approval rate Q4), and groups with faster response time (response time Q4) spent more time in their communities following the adoption of post approvals (\textit{Leader TS}: 0.15; 0.10; and 0.15 SDs).
These trends were gradual across quartiles, and contrast with the overall null effect reported in \Figref{fig:posts4}.
In other words, the moderation burden after enabling post approvals depends on the kind of group and on how community leaders proactively moderate the community.

\xhdr{User activity} 
Larger groups experienced larger decreases in the number of posts (\eg, Q4: $-0.79$ SDs) and larger increases in relative activity (\eg, 0.22 SDs for \textit{Time spent} per post).
Groups with a higher approval rate (Q3/Q4) experienced smaller decreases in the number of posts and smaller increases in relative activity. The higher the approval rate, the smaller the deviations were from PA-OFF matched communities. To a lesser extent, this was also observed for response time: the faster the response time, the smaller the deviations were.
These results suggest that activity-related changes were greater in larger communities and that the strictness and speed of community leaders in approving or rejecting posts mediated changes in user activity.

\xhdr{Moderation} 
Regardless of group size, response time, or approval rate, the number of posts reported and deleted decreased significantly across quartiles in all three analyses.
Other moderation-related metrics such as members removed also decreased in most cases.
Overall, this suggests that the decrease in potentially problematic content following the enabling of post approvals is robust and that it holds even when groups are very large (\eg $-0.11$ SDs for \emph{Posts reported} per post for group size Q4) or when a vast majority of posts are approved (\eg $-0.15$ SDs for approval rate Q4).
In \Figref{fig:posts1}, we saw that post approvals changed the number of posts through both behavior change and through the filtering done by community leaders. 
Here, we see evidence that, regardless of the strictness of this filtering, the number of posts reported and deleted (normalized per post) decreases.

\section{Discussion and Conclusion}

In this work, we presented a large study of post approvals in Facebook Groups, examining both their adoption and their subsequent impact.
Post approvals was adopted after changes in the groups' dynamics in the weeks prior: user activity and moderation increased in the weeks before the setting is enabled, and, on the day when the setting was turned on, there was often a surge in moderation activity.
After the setting is adopted, communities become, on average, centered around fewer posts that receive more comments and reactions, and which users interact with for longer.
These posts were less likely to be reported, and members were less likely to be removed, which suggests an increase in the quality of the discussions happening in the group.
However, the strength of these effects varied with group size and with how proactive moderation was carried out -- \eg in larger groups, leaders spent more time in the group after the setting was enabled, while in smaller groups, they spent less time.
Overall, the findings provide preliminary insight on how proactive moderation may improve online information ecosystems: by adding participation friction in online communities, post approvals elicit behavior change (\cf \Figref{fig:posts1}).

A limitation of our work is that we focus on a limited set of community-level analyses over a short period without considering spillover effects.
This limitation suggests several potential extensions.
First, future work could analyze how post approvals impact the participation of different kinds of users (including leaders); for instance, the setting may disproportionately affect highly-active users
or may affect newcomers more than veteran members of a group, discouraging the former from participating.
Related, future work could examine the reasons for the decrease in post attempts -- to what extent do post approvals discourage lower-quality posts vs. all posts?
Second, future work could investigate the impact of post approvals in the long run. In other words, do the patterns we observe here continue for months or even years? How do communities evolve with and without the setting?
Third, future work might examine spillover effects across different communities. If two communities have many overlapping members and one adopts post approvals (as well as stricter moderation practices), does this influence the behavior of users in the other community which did not adopt the setting?
Fourth, future work could explore other community-level variables, such as within-group friendship network properties. This work may include examining if these variables can help explain the adoption of post approvals (as in \Secref{sec:rq1}), if they change suddenly after post approvals are enabled (\Secref{sec:rq2}), or if the effect of post approvals is heterogeneous across these variables (\Secref{sec:het}).

Last, we note that our analysis was limited to Facebook Groups. 
Adapting the methodology here to explore participation controls studied qualitatively in other platforms such as ``chat modes'' in live streaming platforms~\cite{seering2017shaping} or software such as Reddit's \textit{AutoModerator}~\cite{jhaver2019human} remains future work.
As argued by \citet{kraut2012building}, participation controls are ``design levers'' that shape how people connect with others in online communities.
Thus, understanding how they work may inform the design of better-governed online spaces.

\appendix

\section{Propensity Score Matching}
\label{app:psm}

Throughout the paper, we performed one-to-one propensity score matching (PSM) of PA-ON and PA-OFF communities on the group-level variables. 
We considered all time-varying variables in \Tabref{tab:cov} (under the headers "activity-related" and "moderation-related") as well as all time-invariant variables (under the header "group characteristics"), with the exception of \textit{Group categories}. We did not match on the latter, but found good covariate balance nonetheless, \cf \Appref{app:topics}.
Matching was done using nearest-neighbor matching without replacement, as implemented in \citet{matchit}.
Propensity scores were obtained using a Gradient Boosting Classifier, as implemented in \citet{pedregosa2011scikit}.
For both matching procedures and for all continuous variables, we obtained absolute standardized mean differences (SMDs) smaller than the commonly-used 0.1 threshold that indicates imbalance~\cite{austin2011introduction}.

\xhdrNoPeriod{What leads to the adoption of post approvals?}
For the analyses done in \Secref{sec:rq1}, we performed PSM using an absolute caliper of $0.5$, discarding 125 PA-ON groups (1.4\%) for which we were not able to find good matches. 
For time-varying variables (under the headers "activity-related" and "moderation-related" in \Tabref{tab:cov}), we considered three days of interest (days $-22$, $-25$, and $-28$). 
In other words, each covariate-day pair corresponds to a distinct feature used by the classifier to obtain the propensity scores (\eg posts on day $-22$, posts on day $-25$, and posts on day $-28$).
Day 0 was when the intervention took place for PA-ON group (\ie, when they enabled the post approvals setting), and it was chosen at random for PA-OFF groups.
Covariate balance for this matching is shown in \Figref{fig:matching_pred}.

\begin{figure}
    \centering
    \includegraphics[width=\linewidth]{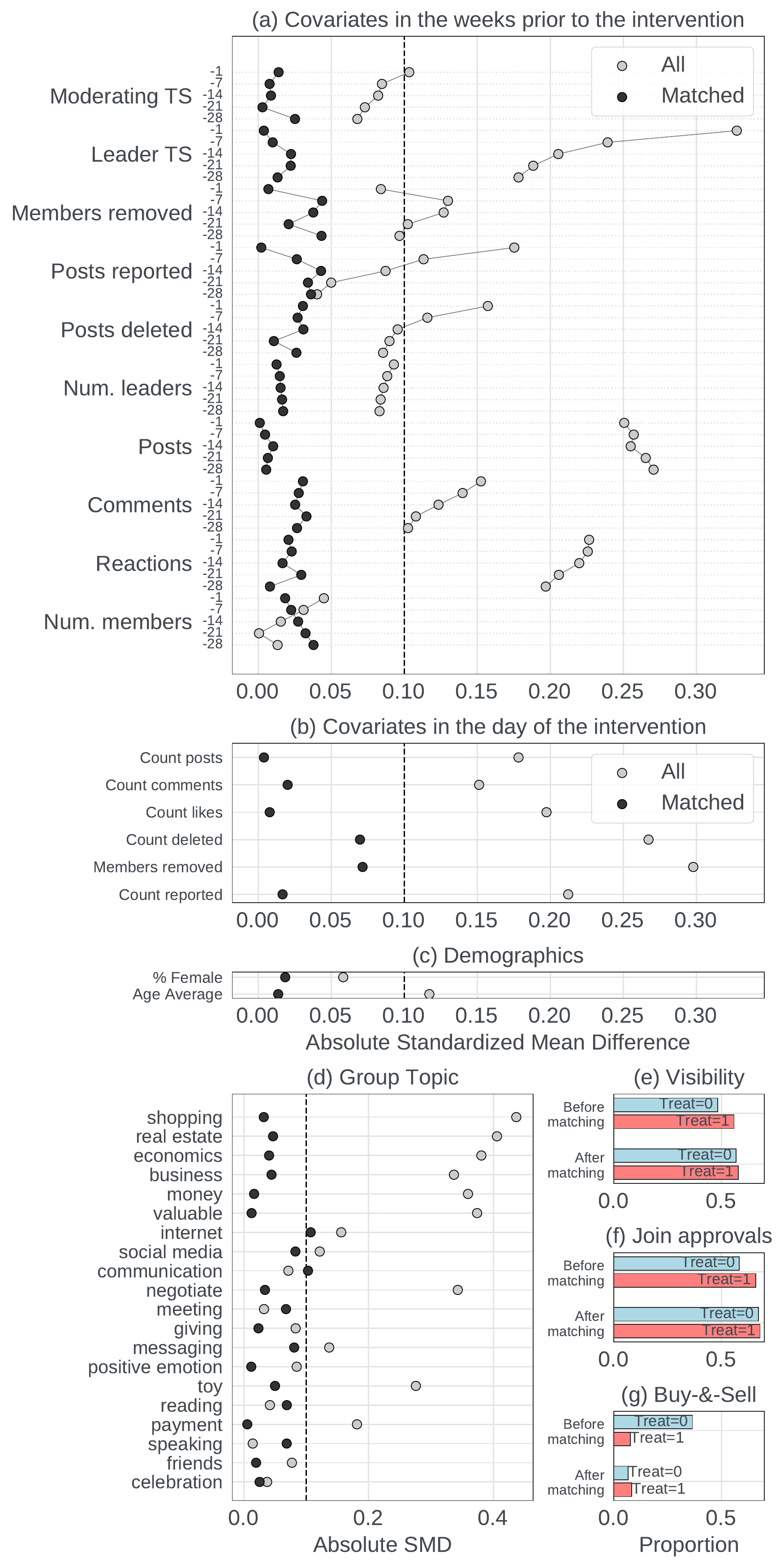}
    \caption{
    Covariate balance for matching done in \Secref{sec:rq2}. 
    \textit{(a)} Absolute standardized mean difference (SMD) for all time-varying variables considered in days $-1$, $-7$, $-14$, $-21$, and $-28$. 
    \textit{(b)} SMD for variables measured in the day of the intervention.
    \textit{(c)} SMD for demographic-related variables.
    \textit{(d)} SMD for the top 20 most popular group categories (from Empath).
    \textit{(e-g)} covariate balance pre\hyp{} and post\hyp{}matching for the three binary variables considered in the matching.}
    \label{fig:matching}

\end{figure}

\xhdrNoPeriod{How do post approvals shape online communities?} 
For the analyses done in \Secref{sec:rq2} and \Secref{sec:het}, we performed PSM with a caliper of $0.0075$, discarding 1426 PA-ON groups (16\%) for which we were not able to find good matches. 
For time-varying variables (under the headers "activity-related" and "moderation-related" in \Tabref{tab:cov}), we considered five days of interest (days $-28$, $-21$, $-14$, $-7$ and $-1$).
Additionally, we considered the number of posts, comments, reactions, deleted posts, reported posts and members removed on the day of the intervention (day 0) measured up to the hour when the intervention (or pseudo-intervention) was introduced. These were all variables that we were able to measure hourly.
Matching on the day of the intervention was done on an hourly basis as PA-ON communities have periods of activity with and without the post approval setting.
Covariate balance for this matching is shown in \Figref{fig:matching}.

\xhdr{Additional observations}  We clarify a couple of decisions regarding the propensity score matching procedure:

\begin{itemize}
    \item As mentioned in \Secref{sec:data}, to obtain candidate PA-OFF groups for matching, we used a random sample of 50\% of all communities that had over 128 members and at least one post and one comment over any 7-day period. This reduces the matching space, as we could have used 100\% of all communities. Yet, empirically, using more than 50\% of the sample harmed the propensity matching score matching capacity to balance the variables of interest as the class imbalance was too extreme. Even with a 50\% sample, before matching, the number of PA-OFF groups ($n=224{,}635$) outnumbered PA-ON groups ($n=8{,}767$) by approximately 25 to 1.
    \item Differences between the PSM done for \Secref{sec:rq1} \vs for \Secref{sec:rq2} and \Secref{sec:het} are as follows:
    \begin{itemize}
        \item Different calipers were used, as achieving covariate balance was harder for the matching in Sec.~\ref{sec:rq2}/\ref{sec:het}.
        \item Different dates were considered for time-varying variables ($-22$, $-25$, $-28$ \vs $-1$, $-7$, $-14$, $-21$, $-28$).
        \item The matching used in Sections~\ref{sec:rq2} and \ref{sec:het} additionally included variables from the day when post approvals was turned on, measured up to the very hour when the setting was changed (\cf \Figref{fig:matching}b)
    \end{itemize}

\end{itemize}

\section{Group Topics}

\label{app:topics}

\begin{figure}
    \centering
    \includegraphics[width=\linewidth]{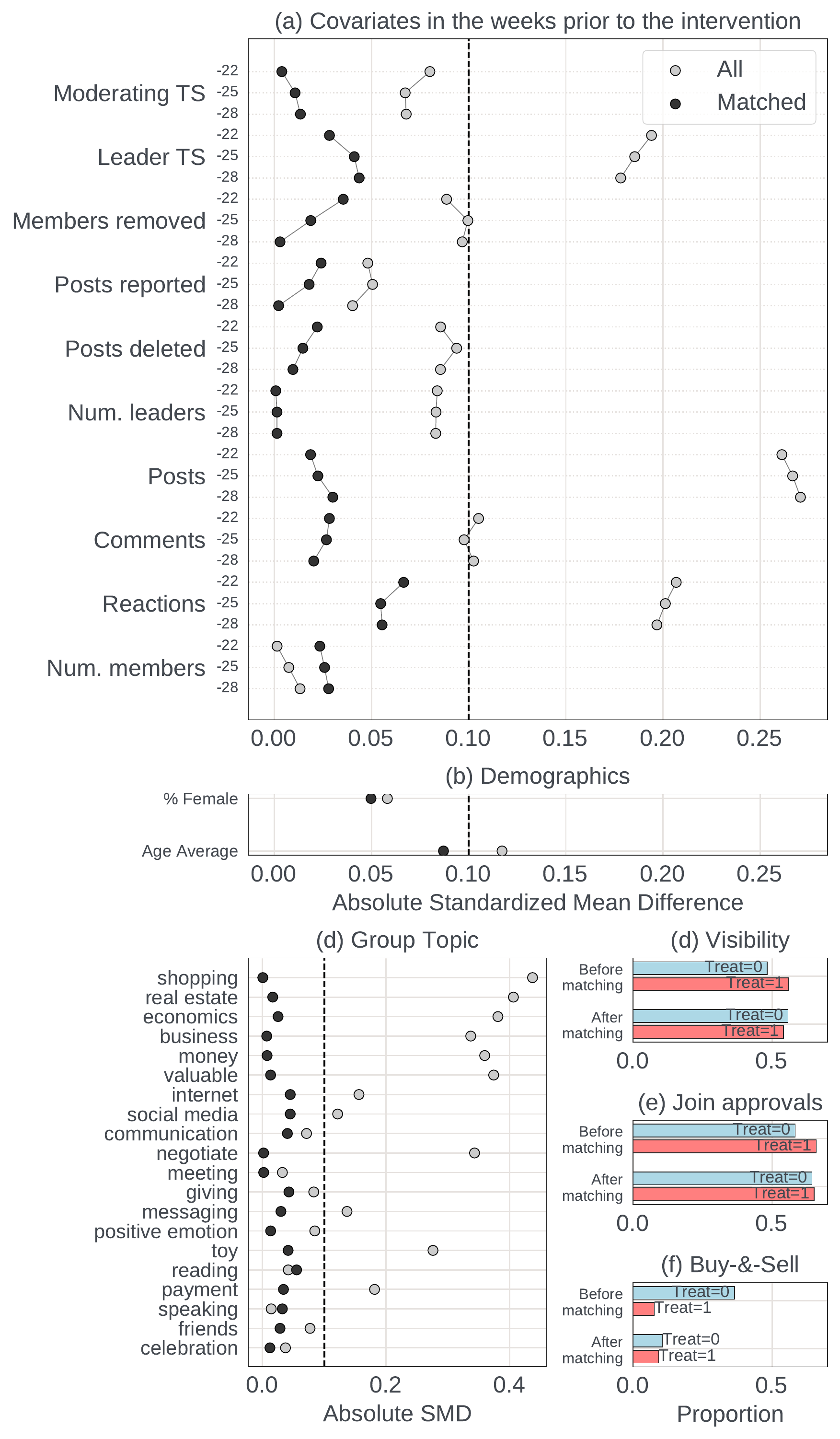}
    \caption{
    Covariate balance for matching done in \Secref{sec:rq1}. 
    \textit{(a)} Absolute standardized mean difference (SMD) for all time-varying variables considered in days $-28$, $-25$ and $-22$. 
    \textit{(b)} SMD for demographic-related variables.
    \textit{(c)} SMD for the top 20 most popular group categories (from Empath).
    \textit{(d-f)} covariate balance pre\hyp{} and post\hyp{}matching for the three binary variables considered in the matching. 
    }
    \label{fig:matching_pred}
\end{figure}

To ensure that PA-ON and PA-OFF groups were topically comparable after propensity score matching, we used Empath~\cite{fast2016empath}'s lexical categories.
Lexicons have been used in causal inference by \citet{saha2019social} and by \citet{sridhar2019estimating}.
We translated group titles and descriptions into English and measured the occurrence of words matching each of the 194 default Empath categories (e.g., \emph{work}, \emph{celebration}, \textit{writing}, \etc).
Even without explicitly matching groups by Empath category word frequency, propensity score matching yielded good covariate balance -- the standardized mean difference for all 194 categories was below 0.1,%
\footnote{Except for the categories \emph{communication} and \emph{internet} in the PSM done for \Secref{sec:rq2}, where they had SMDs equals to 0.103/0.107.} suggesting the two sets of groups had similar titles and descriptions.
\Figref{fig:matching_pred}c and \Figref{fig:matching}d show the covariate balance of the top 20 most common Empath categories before and after PSM.

\section{Additional Plots}
\label{app:adplots}
\begin{figure}
    \centering
    \includegraphics[width=\linewidth]{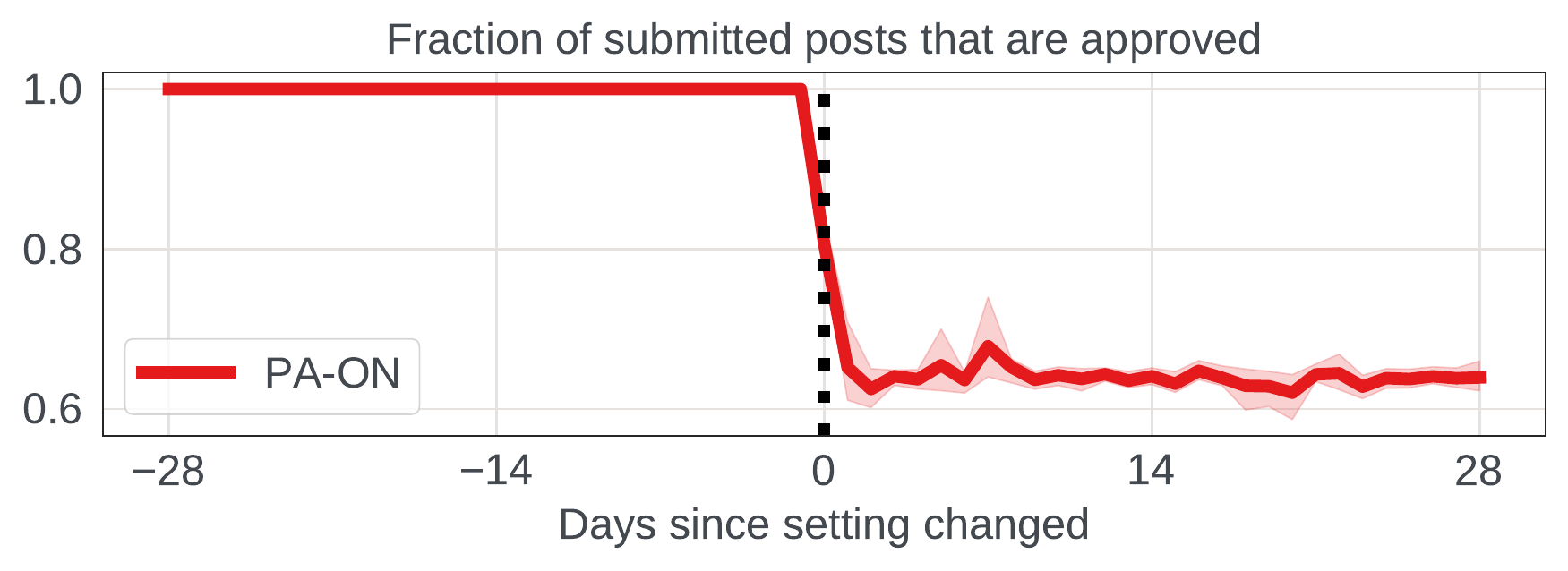}
    \caption{
    For PA-ON groups, we show the fraction of posts submitted that were approved.
    }
    \label{fig:pct}
\end{figure}

\begin{figure}
    \centering
    \includegraphics[width=\linewidth]{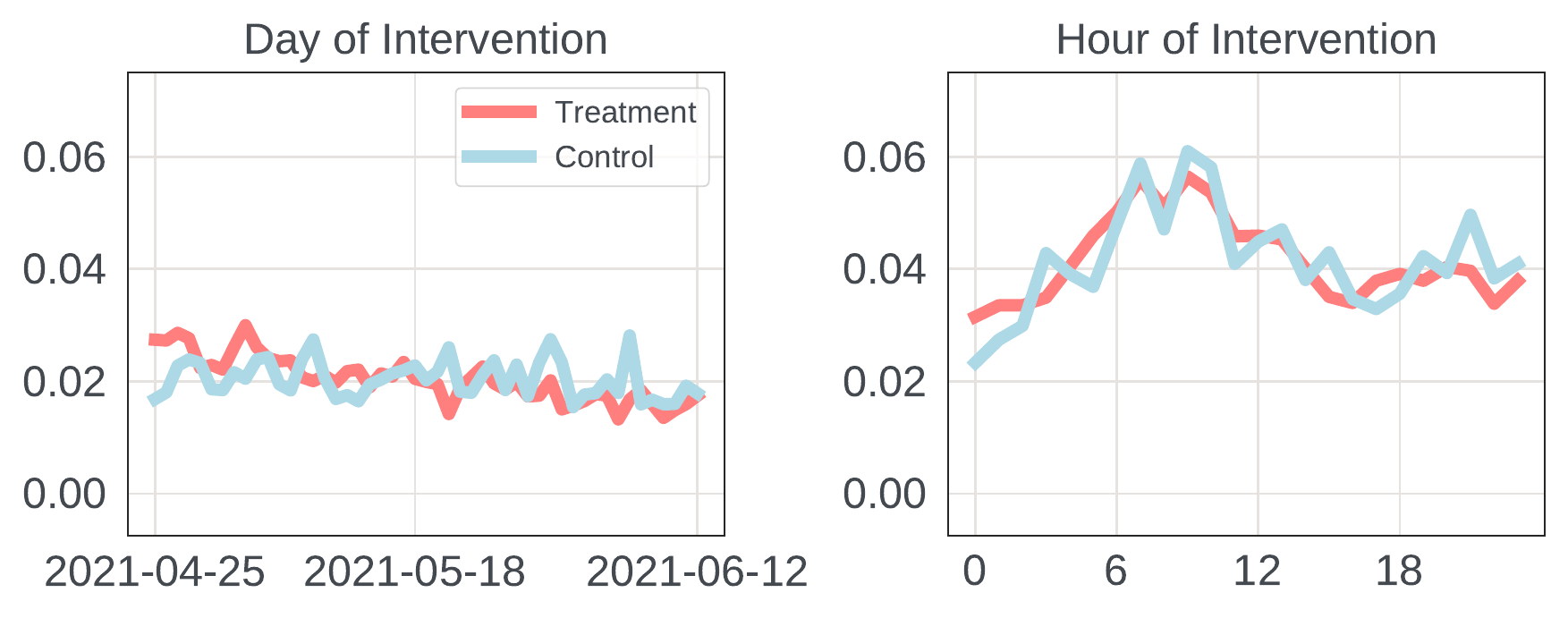}
    \caption{We show the day (left) and hour (right) of the intervention (for PA-ON groups) and pseudo-intervention (for PA-OFF groups) after the matching done in Sections~\ref{sec:rq2} and \ref{sec:het}. Note that the hours are shown in UTC+0.}
    \label{fig:day_hour}
\end{figure}

\begin{figure}
    \centering
    \includegraphics[width=\linewidth]{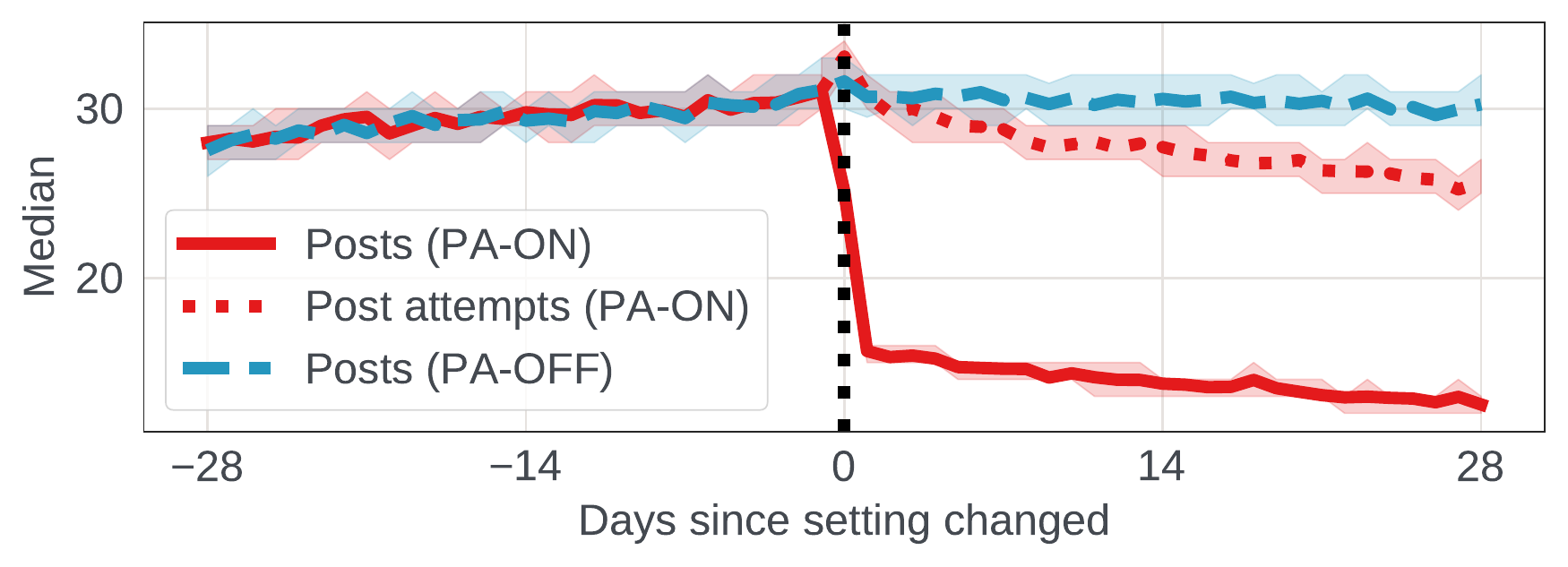}
    \caption{This figure repeats the analysis done in \Figref{fig:posts1} using the median instead of the winsorized mean.}
    \label{fig:median}
\end{figure}

We provide a couple of additional plots as sanity checks:

\begin{itemize}
    \item \Figref{fig:day_hour} shows the distribution of hour and day of the intervention for the matching done for Sections~\ref{sec:rq2} and \ref{sec:het}.
    \item \Figref{fig:median} reproduces \Figref{fig:posts1} using the median instead of the (winsorized) mean.
    \item Complementing \Figref{fig:posts1}, \Figref{fig:pct} shows the fraction of posts submitted that were approved.

\end{itemize}

\FloatBarrier

{
\bibliography{refs}
}

\end{document}